\documentclass[a4paper,11pt]{article}
\usepackage{pos}
\usepackage{physics}
\usepackage{dsfont}

\usepackage{xspace}
\newcommand{\QCED}{QCD$\,+\,$QED\xspace}

\title{Thermodynamics of \QCED from the lattice}

\author[a]{G. Endr\H{o}di}
\author*[a]{E. Garnacho-Velasco}
\author[a]{G. Mark\'{o}}
\author[a]{L. Peth\H{o}}

\affiliation[a]{Institute of Physics and Astronomy,
ELTE E\"otv\"os Lor\'and University,\\
P\'azm\'any P.\ s\'et\'any 1/A, H-1117 Budapest, Hungary}  

\emailAdd{gergely.endrodi@ttk.elte.hu}
\emailAdd{egarnacho@bodri.elte.hu}
\emailAdd{marko.gergely@ttk.elte.hu}
\emailAdd{petho.levente@ttk.elte.hu}

\abstract{QED corrections to observables in QCD are by now well-understood at zero temperature, where QED effects are perturbative and required for precision physics observables, such as the neutron-proton mass difference or the muon anomalous magnetic moment, among many others. However, there are indications in the literature that at finite temperature the introduction of QED can lead to non-perturbative effects, affecting the structure of the phase diagram, since the combined theory is exactly center-symmetric. In this proceedings article, we analyze the structure of the combined theory and the different strategies to simulate it on the lattice, as well as presenting novel results of lattice \QCED simulations at finite temperature, focusing on understanding the role of the electromagnetic coupling in the thermodynamics of this system. We also discuss new features that arise in the combined theory and the challenges that these simulations entail, which require the development of new algorithms in order to understand the phase diagram of \QCED from first principles.}

\FullConference{The 43rd International Symposium on Lattice Field Theory (Lattice 2026)\\
July 26 to August 1, 2026\\
University of Maryland, College Park, USA\\}

\begin{document}
\maketitle

\section{Introduction}
The study of QED corrections at zero temperature has become increasingly relevant in the lattice QCD community, where it has been used to calculate a wide variety of observables like hadron-mass splittings, decay rates of heavy mesons or the anomalous magnetic moment of the muon, see Ref.~\cite{Patella:2017fgk} for a review on these applications. These studies are motivated by the need of the theory to match the increasing precision achieved in the current era of high-energy physics experiments. Although the inclusion of QED in lattice QCD simulations is not exempt from challenges, especially due to finite volume effects~\cite{Patella:2017fgk}, the state-of-the-art formalisms developed in recent years allow for full dynamical \QCED simulations at physical quark masses, which have had an important impact on our current understanding of the Standard Model, see e.g.~the muon anomaly~\cite{Borsanyi:2020mff}.

However, much less is known about QED effects in the thermodynamics of strong interactions. There is some evidence that QED could have a non-perturbative effect on the finite-temperature physics of QCD, owing to the fact that the combined theory of \QCED exhibits center symmetry, also present in a pure gluonic theory and broken by the inclusion of fermions in QCD. This possibility was studied in a series of preliminary works in two-color QCD at strong QED coupling~\cite{Edwards:2010ew}, where it was found that the inclusion of QED turns the deconfinement transition into a real phase transition, similar to the one in pure SU(2) theory. In this work, we will explore the thermodynamics of \QCED using physical quark masses and QED couplings, to understand how the inclusion of dynamical photons affects various quantities relevant at finite temperature.

\section{Center symmetry}

We discuss now the center symmetry of \QCED, for which the partition function is given by 
\begin{equation}
    \mathcal{Z}=\int \mathcal{D}U\,\mathcal{D}\varphi\, \prod_{f=u,d,s}\det M_f[U,\varphi,q_f]\, e^{-S_g-S_\gamma}\,,
\end{equation}
with $S_g$ the gluonic action (in any center-symmetric discretization), $S_\gamma$ the compact photonic action
\begin{equation}
\label{eq:photon_action}
    S_\gamma=\beta_{\rm em}\sum_{n}\sum_{\mu,\nu,\mu<\nu}\qty(1-\cos\qty[\varphi_\mu(n)+\varphi_\nu(n+\hat{\mu}) -\varphi_\mu(n+\hat{\nu}) -\varphi_\nu(n)])\,,
\end{equation}
with the bare coupling $\beta_{\rm em}=1/e^2$ and $M$ the staggered Dirac operator
\begin{equation}
\label{eq:stag_action}
    M^f_{n,l}=\dfrac{1}{2}\sum_{\mu}\eta_\mu(n)\qty[U_\mu(n)u_{f,\mu}(n)\delta_{n,l+\hat{\mu}}-U^\dagger_\mu(n)u_{f,\mu}^*(n)\delta_{n,l+\hat{\mu}}]+m_f \delta_{n,l},
\end{equation}
where the U(1) links are $u_{f,\mu}(n)=e^{iq_f \varphi_\mu(n)}$. Notice we consider these particular actions for concreteness, but the symmetry considerations are independent of the choice for the fermion discretization or the compact or non-compact version of QED.

Pure SU(3) Yang-Mills theory is invariant under $Z_3$ center symmetry transformations
\begin{equation}
     U_4(\vec{n},N_t-1)\rightarrow z U_4(\vec{n},N_t-1)\quad  \forall \vec n\qquad\text{with} \quad z\in Z_3=\{1,e^{i2\pi/3},e^{-i2\pi/3}\}\,\mathds{1}\,,
     \label{eq:z3_trafo_qcd}
\end{equation}
a symmetry that is spontaneously broken at high temperatures, signaling the deconfinement transition. In QCD, the introduction of fermions breaks this symmetry explicitly, since the center transformation is incompatible with the antiperiodic temporal boundary conditions of fermions, see e.g. Ref.~\cite{Fukushima:2017csk} for a detailed analysis. Equivalently, the fermion determinant is not invariant under the transformation~\eqref{eq:z3_trafo_qcd}, breaking center symmetry explicitly. However, this symmetry is recovered in \QCED because the center rotation of the quarks can be compensated with a U(1) rotation of the photon field. In terms of the determinant, it is readily checked that the determinant containing the Dirac operator~\eqref{eq:stag_action} is invariant under the following combined global transformation, 
\begin{align}
    &U_4(\vec{n},N_t-1)\rightarrow z U_\mu(\vec{n},N_t-1)\quad\forall \vec n\,,\\
    \label{eq:z3_trafo_gamma}
    &\varphi_4(\vec{n},N_t-1)\rightarrow \varphi_4(\vec{n},N_t-1)+2\pi\quad\forall \vec n\,,
\end{align}
where, without loss of generality, we consider $z=\tilde{z}\,\mathds{1}=e^{i2\pi/3}\,\mathds{1}$ $z\in Z_3$. 

Then the photon links transform as
\begin{align}
    &u_{u,4}(\vec{n},N_t-1) \rightarrow  u_{u,4}(\vec{n},N_t-1)\,e^{i4\pi/3}=u_{u,4}(\vec{n},N_t-1)\, \tilde{z}^{-1}\,,\\
      &u_{d/s,4}(\vec{n},N_t-1) \rightarrow  u_{d/s,4}(\vec{n},N_t-1)\,e^{-i2\pi/3}=u_{d/s,4}(\vec{n},N_t-1) \,\tilde{z}^{-1}\,,
\end{align}
canceling the center transformation of the gluons. Notice that the transformation~\eqref{eq:z3_trafo_gamma} also leaves the photon action~\eqref{eq:photon_action} invariant.

The recovery of center symmetry could have important implications for the deconfinement transition. In the case of pure SU(3), which is center symmetric, this transition is first-order~\cite{Boyd:1996bx}, while in QCD, where quarks break the symmetry, it is a smooth crossover~\cite{Borsanyi:2010bp}. As we discussed above, Refs.~\cite{Edwards:2010ew} obtained preliminary indications that the inclusion of strongly-coupled QED in two-color QCD, restoring center symmetry, turns the deconfinement transition into a second-order phase transition, the same type of transition as a pure SU(2) theory. 
Note moreover that in QCD, center symmetry can also be recovered for a special choice of the imaginary isospin chemical potential $\mu_I=i2\pi/3$ for three mass-degenerate quarks. Recent lattice simulations showed that this theory exhibits a first-order deconfinement phase transition~\cite{Endrodi:2026tqa}. It is then a natural question to ask how the thermodynamics of physical \QCED behave in comparison to a pure SU(3) theory and QCD.

\section{Lattice setup}
We use $2+1$ flavors of rooted staggered fermions at the physical point\footnote{The LCP does not contain QED effects, so we consider the isospin-symmetric physical point in QCD.} with two steps of stout smearing in the SU(3) links, while for the gluon sector we use the tree-level improved Symanzik action. We will consider both the compact version of QED and the non-compact version in Feynman gauge with Fourier acceleration in the HMC, following Ref.~\cite{BMW:2014pzb}.

In QCD$\,+\,$QED, we can consider different types of Polyakov loops that affect the determinant. First, we have the usual smeared SU(3) Polyakov loop
\begin{equation}
P=\dfrac{1}{V}\sum_{\vec{n}}\Tr\prod_{n_t=0}^{N_t-1}U^F_4(\vec{n},n_t)\,
\end{equation}
with $U^F_4(n)$ being the smeared SU(3) links. We also have the U(1) Polyakov loops, which depend on the electric charge of the flavor
\begin{equation}
    p_f=\dfrac{1}{V}\sum_{\vec{n}}\prod_{n_t=0}^{N_t-1}u_{f,4}(\vec{n},n_t)\,.
\end{equation}
We will analyze the interplay between these two observables at finite temperature in \QCED.

Defining QED in a finite volume entails several difficulties due to the existence of an exact photon zero mode, since QED does not have a mass gap at physical weak coupling, see Ref.~\cite{Patella:2017fgk} for a detailed review of this issue. A well-known consequence is that the Coulomb law cannot be fulfilled in a finite periodic volume for a non-zero electric charge, forbidding the definition of charged propagators. There are several ways of addressing this problem, for example by introducing a photon mass as a regulator~\cite{Clark:2022wjy}, or considering $C^*$-boundary conditions~\cite{RCstar:2022yjz}. Another possibility is to remove this photonic zero-mode from the path integral, which can be done in different ways. Some commonly used strategies are  
\begin{itemize}
    \item QED$_{TL}$: see e.g.~\cite{Blum:2007cy}, this QED version violates reflection positivity~\cite{BMW:2014pzb}
    \begin{equation}
        \sum_{n}A_\mu(n)=0 \quad\forall \mu \quad \Rightarrow \quad \widetilde{A}_\mu(\vec{k}=\vec{0},k_4=0)=0\quad\forall \mu \,.
    \end{equation}
        \item QED$_{L}$: introduced in Ref.~\cite{Hayakawa:2008an}
    \begin{equation}
        \sum_{\vec{n}}A_\mu(\vec{n},t)=0 \quad\forall \mu,t \quad \Rightarrow \quad \widetilde{A}_\mu(\vec{k}=\vec{0},k_4)=0\quad\forall \mu,k_4\,.
    \end{equation}  
\end{itemize}
Another recently proposed scheme is QED$_{r}$~\cite{DiCarlo:2025uyj}. At finite temperature, all these different ways of handling the photonic zero-mode break center symmetry explicitly. Therefore, in analogy with QED$_{L}$,  we propose a novel way of removing the zero-mode at finite temperature which we refer to as QED$_{\mu L}$: 
\begin{equation}
 \widetilde{A}_\mu(\vec{k}=\vec{0},k_4)=0\quad\forall \mu,k_4 \quad \text{except}\quad \mu=4 \,\,\,\text{and}\,\,\,k_4=0\,,
\end{equation}
maintaining the exact center symmetry of the theory.
The idea is that at finite temperature the photon obtains a thermal mass, which manifests as a mass gap for $\widetilde{A}_4(k=0)$, protecting the theory from infrared divergences coming from this constant mode at finite $T$. The fluctuations of this particular mode, which can be understood as a global charge chemical potential $\mu$, are precisely the ones that allow the existence of the center symmetry of the theory and the tunneling between SU(3) sectors. 

\begin{figure}[t]
 \centering
 \includegraphics[width=1.0\linewidth]{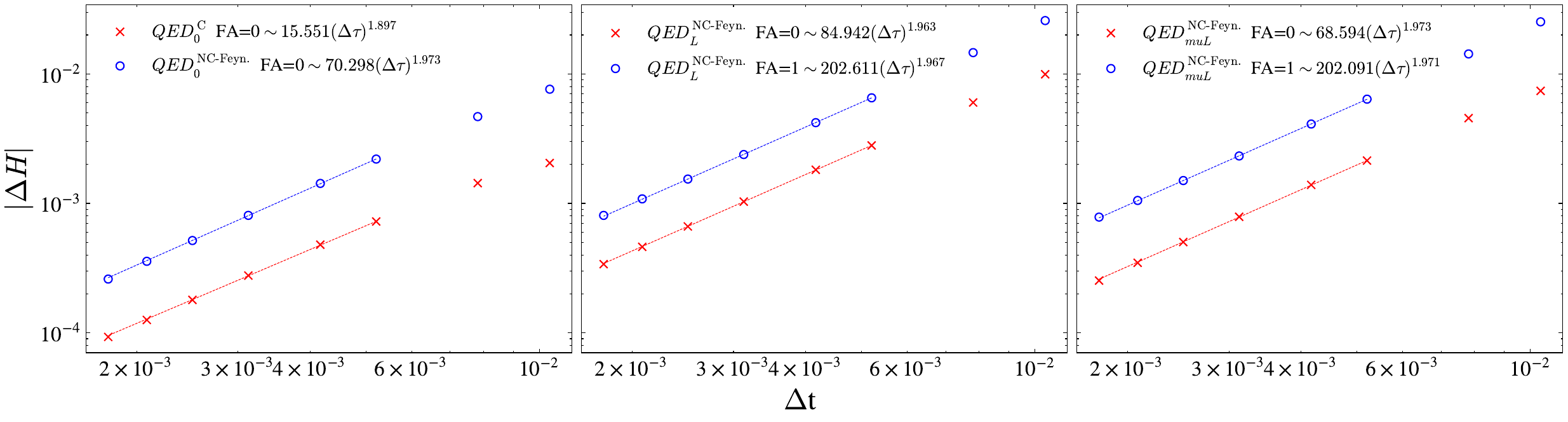}
 \caption{Hamiltonian violation $\Delta H$ as a function of the molecular dynamics integration step-size $\Delta t$. The subscript $0$ denotes no zero-mode removal and FA refers to the Fourier accelerated HMC.}
 \label{fig:deltaH}
\end{figure}

\begin{figure}[b]
 \centering
 \includegraphics[width=0.85\linewidth]{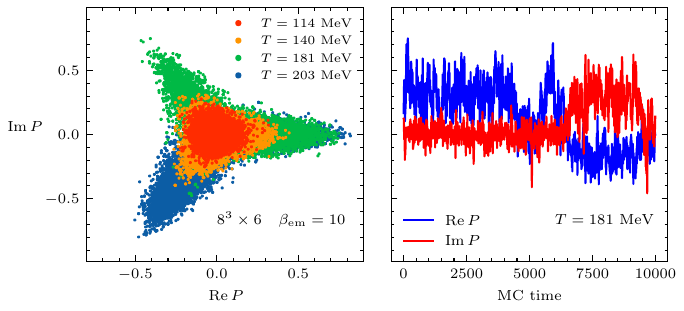}
 \caption{Scatter plot of the SU(3) Polyakov loop $P$ on an $8^3\times6$ ensemble with $\beta_{\rm em}=10$, for different temperatures (left) and Monte Carlo time history of the real and imaginary parts of $P$ at  $T=181$ MeV (right).}
 \label{fig:ploop_scatter_compact}
\end{figure}

\begin{figure}[t]
 \centering
 \includegraphics[width=0.6\linewidth]{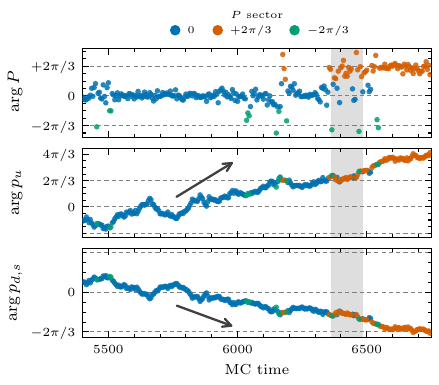}
 \caption{Illustration of the tunneling between center sectors in a $8^3\times6$ lattice with $\beta_{\rm em}=10$ and $T=181$ MeV. The color corresponds to the SU(3) Polyakov loop sector.}
 \label{fig:ploop_tunnel}
\end{figure}

\section{Results}
We present now some results of these novel \QCED simulations at finite temperature, considering different setups. First of all, in Fig.~\ref{fig:deltaH}, we show a crosscheck of our \QCED code, where the Hamiltonian violations $\Delta H$ depend exponentially on the molecular dynamics step-size time $\Delta t$ as expected, for all the QED actions we have considered.

\subsection{Compact QED and electromagnetic coupling scan}
Next we focus on the compact QED formalism with no zero-mode removal. In the left panel of Fig.~\ref{fig:ploop_scatter_compact} we show a scatter plot of the SU(3) Polyakov loop at different temperatures, in a $8^3\times6$ lattice and weak QED coupling $\beta_{\rm em}=10$, below and above the transition. We can see that at high $T$, center symmetry is spontaneously broken, signaled by non-zero values of $P$ on typical configurations, just as it occurs in QCD. However, the existence of center symmetry in \QCED allows the Polyakov loop to explore the three center sectors above $T_c$. These transitions between different sectors are visualized in the right panel of Fig.~\ref{fig:ploop_scatter_compact}. Even in this small volume, the transitions are found to be rare. 

In Fig.~\ref{fig:ploop_tunnel}, we show how the tunneling mechanism works, where the U(1) Polyakov loops $p_u$ and $p_{d/s}$ rotate (in opposite directions due to the electric charges) until they reach the center element value that cancels the SU(3) one, allowing for the SU(3) Polyakov loop to change from the real to the positive imaginary sector. This also explains why the tunneling processes are rare, 
since the HMC algorithm used for sampling the configurations 
only proposes small changes to the link variables, making 
global rotations of the U(1) links unlikely to occur. 
We observed that this is increasingly so the case as the volume grows,
rendering the study of the order parameter distribution near $T_c$
challenging. We are currently working on developing algorithms to help improve the ergodicity of the sampling of the theory, like the introduction of local center updates between HMC trajectories.

\begin{figure}[b]
    \centering
    \includegraphics[width=0.6\linewidth]{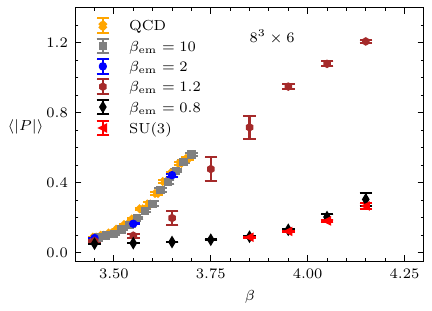}
    \caption{SU(3) Polyakov loop $P$ as a function of the QCD inverse coupling for different values of $\beta_{\rm em}$, as well as in a pure SU(3) theory and in QCD.}
    \label{fig:ploop_bemdep}
\end{figure}

Finally, we also scanned the theory in $\beta_{\rm em}$ space, going from the weakly-coupled (physical) regime of QED to the strongly coupled one, in order to compare with the setup used in two-color QCD in Refs.~\cite{Edwards:2010ew}. In Fig.~\ref{fig:ploop_bemdep}, we show the results for the SU(3) Polyakov loop as a function of the QCD inverse coupling $\beta$, for different values of $\beta_{\rm em}$. We observed that at weak electromagnetic couplings $\beta_{\rm em}>1$, the results for $P$ agree with the ones in QCD. However, when $\beta_{\rm em}\sim1$ the results depart from the ones in QCD and when $\beta_{\rm em}<1$ they appear to approach the result in pure SU(3). The similarity between the results of the electrically strongly-coupled \QCED theory and pure SU(3) Yang-Mills theory agrees qualitatively with the findings of Refs.~\cite{Edwards:2010ew}. We also conjecture that this behavior could be related to the phase transition of the compact QED lattice model at $\beta\approx1$~\cite{Kogut:1987kg}, although we are currently investigating the precise mechanism behind it.

\subsection{Non-compact QED and zero mode removal}
\begin{figure}[t]
    \centering
    \includegraphics[width=0.6\linewidth]{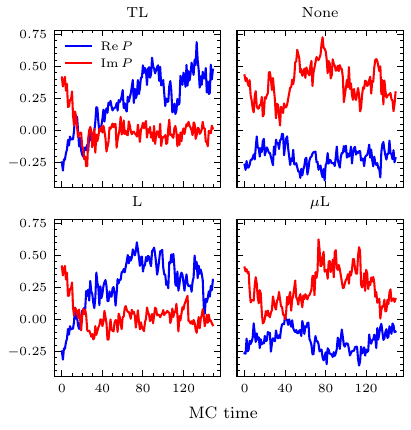}
    \caption{Monte Carlo time history of the real and imaginary parts of the SU(3) Polyakov loop $P$ for different zero-mode subtraction schemes, in an $8^3\times6$ lattice with $\beta_{\rm em}=10$ and $T=187$ MeV.}
    \label{fig:ploop_zero_mode}
\end{figure}

We now present the results using the non-compact QED formulation in Feynman gauge. In Fig.~\ref{fig:ploop_zero_mode}, we demonstrate how the different zero-mode removals behave. We start the simulation from a configuration with the SU(3) Polyakov loop in the positive imaginary sector, and we can see that in the QED$_{TL}$ and QED$_{L}$ formulations, which break center symmetry explicitly, the trajectories drive $P$ into the real sector. However, without removing the zero-mode (None) and in QED$_{\mu L}$, which respect center symmetry, the Polyakov loop can stay exploring the imaginary sectors.

\begin{figure}[t]
    \centering
    \includegraphics[width=0.9\linewidth]{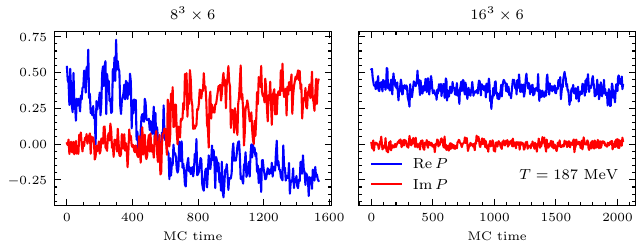}
    \caption{Monte Carlo time history of the real and imaginary parts of the SU(3) Polyakov loop $P$ at $T=181$ MeV and $\beta_{\rm em}=10$, for an $8^3\times6$ lattice (left) and a $16^3\times6$ lattice (right), using QED$_{\mu L}$.}
    \label{fig:ploop_nc}
\end{figure}

Finally, in Fig.~\ref{fig:ploop_nc} we show that the SU(3) Polyakov loop behaves in the same way as in the compact theory, where transitions are rare but present for an $8^3\times6$ lattice, but absent when the volume is larger.

\section{Summary and Outlook}
In this proceedings article, we have discussed the symmetry structure of \QCED at finite temperature, focusing on the role of center symmetry in the combined theory. We have presented novel results on the thermodynamics of physical \QCED, demonstrating the existence of center symmetry and how this allows for the tunneling between different sectors for the SU(3) Polyakov loop at high temperatures. However, sampling the different phases of the theory is challenging even in small volumes, both in the compact and non-compact versions of QED. This ergodicity issue might be alleviated by adapted sampling algorithms. In addition, we have shown that in the strong coupling limit, the results for the SU(3) Polyakov loop in \QCED approach that of pure SU(3) theory, in qualitative agreement with previous preliminary studies in two-color QCD. Finally, we have discussed the required removal of the photonic zero-mode in QED at finite temperature, where usual schemes break center symmetry explicitly, and introduced a new way of addressing this issue that we dubbed QED$_{\mu L}$, which respects this symmetry, and hence is suited for finite-temperature \QCED simulations.

As the next steps, we will investigate the introduction of local updates to improve the ergodicity of the HMC sampling of the theory, as well as to study the approach towards the continuum limit. In addition, we plan to extend our study of the gQED effects on the thermodynamics of QCD to the equation of state and quark number susceptibilities, which play an important role in understanding the phase diagram of strongly-interacting matter.

\acknowledgments
This work was funded by the Hungarian National Research, Development and Innovation Office (Research Grant Hungary 150241) and the European Research Council (Consolidator Grant 101125637 CoStaMM). The authors are grateful for enlightening discussions with K\'{a}lm\'{a}n Szab\'{o} and Lorenz von Smekal. 
We acknowledge the Digital Government Development and Project Management Ltd.\ for awarding us access to the Komondor HPC facility based in Hungary.

\end{document}